\documentclass[prd, preprint, 12pt]{revtex4-1}

\usepackage{amsmath}
\usepackage{amssymb}
\usepackage{setspace}
\usepackage{graphicx}
\usepackage{natbib}
\usepackage{float}

\begin{document}
 
 % ! TEX spellcheck
 %
%\ \vskip 1.0 in

\begin{center}
 {  {\bf Cognitive Science and the Connection between Physics and Mathematics}}

%\smallskip

\vskip 0.1 in

{{\bf Anshu Gupta Mujumdar$^a$ and Tejinder  Singh$^b$}}

\smallskip
{\it $^a$706, Bhaskara, TIFR Housing Complex,}
{\it Homi Bhabha Road, Mumbai 400005, India}\\
{\it $^b$Tata Institute of Fundamental Research,}
{\it Homi Bhabha Road, Mumbai 400005, India}\\
{\tt email: anshusm@gmail.com, tpsingh@tifr.res.in}\\
%\smallskip

%\vskip 0.5cm
\end{center}
\vskip 0.4 in 

\noindent{\it This essay received the Special Prize for Creative Thinking in the 2015 Essay Contest ``Trick or Truth: the Mysterious Connection Between Physics and Mathematics" conducted by the Foundational Questions Institute, USA} [http://fqxi.org] 

\vskip 0.4 in

\centerline{\bf ABSTRACT}

 The human mind is endowed with innate primordial perceptions such as space, distance, motion, change, flow of time, matter. The field of cognitive science argues that the abstract concepts of mathematics are not Platonic, but are built in the brain from these primordial perceptions, using what are known as conceptual metaphors. Known cognitive mechanisms give rise to the extremely precise and logical language of mathematics. Thus all of the vastness of mathematics, with its beautiful theorems, is human mathematics. It resides in the mind, and is not `out there'.

Physics is an experimental science in which results of experiments are described in terms of concrete concepts - these concepts are also built from our primordial perceptions. The goal of theoretical physics is to describe the experimentally observed regularity of the physical world in an unambiguous, precise and logical manner. To do so, the brain resorts to the well-defined abstract concepts which the mind has metaphored from our primordial perceptions. Since both the concrete and the abstract are derived from the primordial, the connection between physics and mathematics is not mysterious, but natural. This connection is established in the human brain, where a small subset of the vast human mathematics is cognitively fitted to describe the regularity of the universe. Theoretical physics should be thought of as a branch of mathematics, whose axioms are motivated by observations of the physical world.

We use the example of quantum theory to demonstrate the all too human nature of the physics-mathematics connection: it is at times frail, and imperfect. Our resistance to take this imperfection sufficiently seriously [since no known experiment violates quantum theory] shows  the fundamental importance of experiments in physics. This is unlike in mathematics, the goal there being to search for logical and elegant relations amongst  abstract concepts which the mind creates.

\smallskip

\setstretch{1.1}

\noindent 

\smallskip

\smallskip

\setstretch{1.4}

\section{Physics: experiments, unification of concepts, and mathematics}

\noindent Mathematics is a precise language in which true statements can be  proved starting from a set of axioms, using logic. [Except for the subtlety posed by 
G\"odel theorems, whose implications for mathematics and physics will not be discussed here]. This provides us with the grand edifice of mathematics, with its beautiful and great theorems, and unification across arithmetic, algebra, geometry, and analysis \cite{Courant}. In mathematics, there does not seem to be a place for experiments. Physics, on the other hand, is an experimental science (hence dependent on technology)  of the world we observe, where experiments couple with great leaps of conceptual unification. The mathematics used in physics comes in only at a later stage, when we seek a precise language to describe the observed physical phenomena.

A brief look at the history of physics will testify to these assertions. Thousands of years ago, physics had qualitative, non-mathematical beginnings in primordial human perceptions of matter, light, shape, pattern recognition, space, time, motion, and {\it elementary\ counting}. [Elementary counting, or number sense, is hard wired into the brain, as we shall see below, and is a pivotal concept common to both physics and mathematics].  

An early observation/experiment of paramount significance had to do with the motion of the sun, the moon, and the planets against the backdrop of the fixed celestial sphere. Elementary knowledge of geometric shapes (such as circles)  was used to describe their assumed motion around the earth. Centuries later, it took a great conceptual leap to assert that our description of nature is simpler if the earth and planets are assumed to go around the sun. This was a historic development which was independent of mathematics. Extraordinary and painstaking observations (made possible by advances in technology) established the shape of the orbit of Mars \cite{Koestler}. Then came the concepts of force and acceleration, followed by one of the greatest unifying principles in physics: the force that causes planets to go around the sun is the same as the force of gravity which causes things on earth to fall to the ground. Mathematics came in when Kepler inferred that the Martian orbit is an ellipse. And when Newton deduced from experiments how force and acceleration are quantitatively related. And how the force of gravity falls off with distance, and how the law of motion combined with the law of gravitation shows that planetary orbits around the sun are elliptical. This  proof has great mathematical  beauty, and the amazing success of such proofs is of course the subject of this discussion. We re-emphasise that in physics, observation/experiment, and concepts and their unification come first, and mathematics makes a later (albeit grand) entry.

Undoubtedly, this character has been repeated throughout the history of physics. Sustained experiments on electricity and magnetism over centuries, and the development of the concept of a field, eventually led to the conceptual understanding that electricity and magnetism are two aspects of the same [electromagnetic] field, to be then followed by the mathematically precise formulation of Maxwell's electrodynamics. The observation that the product of electric and  magnetic permeability equals the inverse square of the speed of light led Maxwell to another leap, namely that light is an electromagnetic wave. The fact that classical physics failed to predict the experimentally observed spectrum of black-body radiation suggested to Planck the novel concept that atoms emit and absorb light in discrete quanta - this led to Planck's radiation formula. The inability of classical physics to explain the features of the photo-electric effect suggested to Einstein that light is made of discrete quanta; then came the precise mathematical relation between  energy and frequency of the photon. Bohr proposed the concept of  quantised angular momentum to explain hydrogen spectra. Finally, Schr\"odinger and Dirac put the work of Planck, Einstein and Bohr on a firm mathematical footing, in their extremely elegant and universal equations. Always, it is experiments and concepts first, and then the mathematical formulation.  

Sometimes, a concept has outlived its time, and must go. The failure of the Michelson-Morley experiment to detect the motion of the earth through the hypothesised ether led Einstein and others to abandon the ether, and look for a set of mathematical coordinate transformations which allow the speed of light to be the same for all inertial observers. Even in the case of general relativity, for which it is said that the theory is built purely on deductive power, one needs to  exercise caution while remarking thus. The roots lie in the remarkable experimental fact that objects of all masses fall in a  gravitational field with the same acceleration, from which comes the conceptual leap that gravitation is space-time curvature. The leap is followed by years of mathematical struggle on Einstein's part, before he arrives at the correct field equations. Even then, scope is left open for whether or not to include the cosmological constant, and the answer has been provided only now (seemingly so), a hundred years later, by astronomical observations!  

On other occasions, one may construct a mathematically consistent theory based on existing experimental knowledge, but the theory is not confirmed by  experiments. A case in point is a relativistic scalar theory of gravitation, which is not viable because it fails to predict the bending of light \cite{MTW}. Or the case of the very elegant and highly symmetric Maxwell equations with magnetic monopoles; to date we do not have experimental evidence for magnetic charges. Hence we do not believe these equations describe nature, even though they are more beautiful than ordinary Maxwell electrodynamics! [Whereas for a mathematician the symmetric equations with monopoles  will be `a thing of beauty is a joy forever' - fun to play with and explore, irrespective of whether or not experiments find monopoles]. 

The mathematics that physicists use is for the most part relatively simple, only a small subset of the vastness of pure mathematics. Often enough, it is first and second order partial differential equations in many variables, some aspects of algebra and group theory [some groups are more special than others], some aspects of geometry (say Euclidean and Riemannian, as opposed to say affine, projective and Riemann-Cartan geometry), and some aspects of real and complex analysis. Number theory rarely makes an appearance. 

One must also note that the mathematics that physics uses is silent about choice of initial conditions! Mathematics describes the laws of physics, which have to be supplemented by initial conditions. What decides the initial conditions of the universe? Are there mathematical laws for them? We do not know, not yet.

We may think of theoretical physics as that subset of pure mathematics whose axioms and concepts are motivated by experiments on the physical world. The consequences of the laws of physics are akin to the theorems that follow from the axioms. Looked at this way, it should not come as a surprise that great physicists sometimes use their equations to predict the outcomes of future experiments. As in Dirac's prediction of anti-matter, and Einstein's prediction of the bending of light. The theoretical law is an all-encompassing mathematically precise  description of phenomena that lie in its domain, motivated by experiment, and predictor of experiments to come, until there comes an experiment which exposes the law's limitation, compelling us to look for a more general law.

\section{Mathematics: Primordial physics, axioms, theorems, and  beauty} 
Remarkably enough, the primordial roots of mathematics are in the same human perceptions as in physics: shape, pattern recognition, counting, space, time, and change. With one significant difference: there is no place in mathematics for matter (material substance), and by extension, for light! This to us is the biggest difference between physics and mathematics, from which all other differences germinate. Matter in mathematics is relevant to the extent that it helps in abstraction, and to arrive at the intuitive notion of a set (of objects). But the kinds of sets more likely to be of interest to a mathematician would be set of integers, set of transcendental numbers, set of all triangles on a plane etc. as opposed to say a physicist's set of planets in the solar system, or set of elementary particles. The commonality of many primordial human perceptions upon which both physics and mathematics build, is to us the reason for the `unreasonable efficacy of mathematics in physics', as we shall argue in the next section.

Abstracting from counting, shape, pattern, and space-time-change, mathematics receives undefined  fundamental entities such as natural numbers, point and line. By giving definitions/axioms, along with operations and  relations  amongst them, followed by generalisations, mathematicians have developed the classic subjects of number theory, algebra, geometry and analysis. Greater unification, still an ongoing program, is endeavoured through developments such as algebraic geometry, algebraic number theory and arithmetic geometry. This enormous edifice of meaningful and beautiful inter-relations between derivatives of the humble primordial abstractions is, at least on the face of it, very different from physics. Even though physicists use a small part of it to describe laws of the physical world.

Perhaps the most striking example of mathematical abstraction and generalisation, and how it is motivated, comes from  the development of the number system. Mathematicians struggled for centuries to understand things which we now teach in high school.  Natural numbers were motivated by objects in the physical world, but are quickly abstracted to entities by themselves, with no reference to physical objects. The fundamental laws of arithmetic (commutativity, associativity, distributivity) govern how we add and multiply numbers. The introduction of the zero and subtraction mark important progress. So does the Indian notational system  for representing integers, and independently, the principle of mathematical induction. The fascinating and beautiful world of prime numbers, the important proof of the prime number theorem, the still unproven Goldbach conjecture that every even number can be expressed as a sum of two primes, and the still unproven statement that there are infinitely many prime pairs $p$ and $p+2$. The only recently proven Fermat's last theorem - being the work of many outstanding mathematicians over centuries. Continued fractions, Diophantine equations, and much more \cite{Courant}.  All this makes us believe that numbers have a life of their own. But where do they live? Subsequently we will try to make the somewhat unconventional case that, despite appearances, they live in the human brain, and nowhere else.   

Negative integers were introduced in order that subtraction of a larger number from a smaller number became meaningful, while defining operations so that the original laws of arithmetic were preserved. Rational numbers were introduced so that the division $b/a$ is meaningful even if the number $b$ is not a multiple of $a$. Once again, the generalisation is carried out ensuring the original axioms are preserved by the larger system, else it would be a futile generalisation. The awe-inspiring incommensurables (irrationals) found their place on the number line as non-repeating infinite decimals. The concept of limit, infinite series, and the real continuum were born, and turned out to be an immense benefit to theoretical physics. The arrival of analytical geometry: every geometric object and operation can be mapped to numbers. The mathematical analysis of the infinite, the denumerability of the rationals, the non-denumerability of the continuum, the cardinality of infinite sets, transfinite numbers, the fascinating undecidable status of Cantor's Continuum Hypothesis [that there is no set whose cardinality is greater than that of the set of integers, but smaller than that of the set of real numbers]. Cohen's non-Cantorian set theory, where the continuum hypothesis is not true. So far one comes, starting from the primordial perception of counting, having built upon the power of meaningful generalisation. And further afield, the world of complex numbers [still obeying the fundamental laws of arithmetic] necessitated by the need to give meaning to solutions of quadratic equations. The theory of functions of a complex variable flourished; yet the Riemann hypothesis remains the most famous unsolved problem in mathematics, and would have important consequences for our knowledge of prime numbers.     

The abstraction of shapes leads to geometry, with no better example than Euclid's historic work in the {\it Elements}, where the modern mathematical method of deduction from axioms was first laid down more than two millennia ago. With further abstraction, we learn that apart from Euclidean geometry, the plane also admits an affine geometry, which preserves straight lines and parallel lines, but not distances; and projective geometry, which does not preserve parallelism \cite{Ruelle}.  Abandoning Euclid's fifth postulate, geometers invented curved geometries, paving the way for Riemann's monumental work on hypersurfaces of arbitrary curvature, and differential geometry. Spaces of arbitrary dimension (including infinite) were introduced.  

Study of shapes independent of metric and projective properties gave birth to the fascinating field of topology. Euler's formula for polyhedra was an early milestone. The elegant and easy to state Four Color Theorem was proved only as recently as 1977 \cite{FourColor}, that too using thousands of hours of computer time, and the proof was later simplified in 1997  and then in 2005. The concept of dimension was generalised as a topological feature. The Hausdorff (fractal) dimension appears intimately in Mandelbrot's theory of fractals. Topological surfaces were classified using concepts such as genus and Euler characteristic. In Falting's theorem, the set of rational points on an algebraic curve make an awesome connection with its genus \cite{Falting}.The fundamental theorem of algebra can be beautifully proved by considerations of a topological character. The study of equivalence of knots received great impetus from the discovery of the Jones polynomial \cite{Jones}, with deep connections to quantum field theory \cite{Witten}. In the study of topology, we once again see how abstraction greatly enriches mathematics.

Pattern recognition, with inputs from arithmetic, is the basis of algebra. The fundamental laws of arithmetic are adapted here, giving rise to the concepts of groups, fields, rings and ideals. Each one of them becomes a branch of mathematics by itself. Just to mention one instance, the classification theorem for finite simple groups is more than 10,000 pages long! Algebraic geometry is the modern incarnation of Cartesian analytical geometry: one studies equations for curves and surfaces in higher dimensions, using complex numbers,  leading to the definition of algebraic varieties (geometric manifestations of solutions of systems of polynomial equations). The subject has led to some of the deepest mathematical developments and new branches of our times. This is the great twentieth century contribution of Serre and Grothendieck, followed by many others.  Arithmetic geometry combines algebraic geometry and number theory.  This is a subfield of algebraic number theory which studies algebraic structures related to algebraic numbers. Profound abstraction is now being followed by profound synthesis and unification in pure mathematics. 

Calculus was of course abstracted from observation of motion and change, and came with the new concept of the infinitesimal. The origins lay in two more immediate problems: to determine the tangent lines of a curve (the birth of differential calculus) and to determine the area within a curve (the birth of integral calculus). The genius of Newton and Leibniz lay in recognising the connection between the two. It is no wonder that calculus, invented to understand motion, revolutionised mechanics and physics. Mathematical concept and nature showed complete harmony, and today physicists employ calculus, more than perhaps any other branch of mathematics, in their work. The theory of nonstandard analysis borrows sophisticated ideas from modern mathematical logic to make infinitesimals `respectable' from a mathematical viewpoint.

A unifying feature of mathematics today is that it can be entirely based on an axiomatic treatment of set theory, and derived from a set of axioms, say ZFC (Zermelo-Fraenkel-Choice). But why these particular axioms; especially why the Axiom of Choice? G\"odel showed that if ZF is consistent, so is ZFC. Some mathematicians do not like the Axiom of Choice, but most consider that accepting the Axiom of Choice yields richer mathematics than if one left it out \cite{Ruelle}. To us, this in itself is a striking demonstration that mathematics is an enterprise of the human mind, and not a universal Platonic truth `out there'. 

Based on hard-wired human {\it primordial\ perceptions} such as object, size, shape, pattern and change, mankind has built {\it abstract} concepts such as numbers, point, line, infinitesimal, infinity, equation, group, curvature, and so many more, on which the vastness of mathematics is built. The same primordial concepts give rise to the {\it concrete} concepts of experimental physics such as force, mass, motion, charge, rotation, and abstract physics/maths  concepts such as field and symmetry. [We do not make a distinction between the abstract concepts of theoretical physics and of mathematics]. Cognitive science, in its application to mathematics, aims to show in a scientific way how the abstract concepts of mathematics build upon primordial perceptions, using what is known as {\it conceptual\ metaphor}. We now argue that when we seek a precise description of inter-relations  amongst the concrete concepts on which experimental physics is based, we necessarily rely on the abstract concepts of mathematics. Since the concrete and abstract are both built upon the primordial, this demystifies the extraordinary success of mathematics in physics. It could not have been otherwise.

\section{Cognitive science, physics, and mathematics}
In their work, physicists and mathematicians generally prefer to ignore or `forget' the brain, treating it as a perfect and passive agent which helps them discover objective truths about the physical and mathematical world out there. But is that not an unscientific stance? Nature did not evolve the brain for doing physics and maths.  How are we ever going to be able to scientifically prove Platonism? Asserting that there is a mathematical universe, which we somehow grasp in an extra-sensory manner, and which perhaps is coincident with the physical universe, is at best an act of faith. Over the last two decades, cognitive neuroscience and cognitive science have made a forceful case that all mathematics is {\it human} mathematics,  made in and by the human brain, and stable across cultures. Whether or not human mathematics coincides with a Platonic mathematics is again not something we can scientifically address. [The coincidence in fact is unlikely, for human mathematics builds on metaphors, whereas Platonism is literal]. Physicists use human mathematics to give a precise description of the regularity they observe in their experiments. No surprises here. The great wonder, which lies beyond all present day physics, mathematics and biology, is the very existence of the physical universe with its regularity, and that too a universe with at least one planet full of intelligent beings! How life and the human brain have evolved to the point where such cognitive mechanisms become operational, is by itself a fundamental question, though beyond the scope of the present discussion.

Cognitive neuroscience, which aims to identify the biological substrates which underlie cognition, including mathematical cognition, has made enormous strides in recent years. Fundamental progress in identifying function specificity of brain regions  has been made possible by advances in brain mapping techniques such as Positron Emission Tomography (PET) and functional Magnetic Resonance Imaging (fMRI). At the same time, classic laboratory experiments \cite{Dehaene} have established that human babies have some innate mathematical abilities. These include, amongst others: ability to discriminate between collections of two and  three items (at age three or four days) \cite{Antell}; one plus one is two (age five months) \cite{Wynn}. These abilities apply to visual as well as auditory stimuli \cite{Bibelac}. Similar abilities have been reported in scientific studies on some animals \cite{Mechner, Church, Hauser}. All human beings, independent of culture and education, are capable of subitizing, this being the ability to instantly tell at a glance whether a collection has one, two or three objects \cite{Dehaene}.There is some evidence that the region of the brain known as inferior parietal cortex is responsible for symbolic numerical abilities \cite{Dehaene}. Remarkably, this is a highly associative area where neural connections from vision, audition and touch also come together. This in itself seems to suggest that primordial physical perceptions (which base on motor-sensory inputs) might relate to abstractions such as numerics. One can thus say that strong evidence has emerged that some very elementary arithmetic abilities are hard wired into the brain: they are neither discovered nor invented, but they pre-exist in neural connections in infants.  [The region in the brain known as prefrontal cortex, which is involved in complex motor routines and planning, also seems to be used in complex arithmetic calculations.  Rote abilities such as memorization of multiplication tables seem associated with basal ganglia]. Cognitive science says that human beings employ ordinary cognitive mechanisms (common with other aspects of understanding and not restricted to mathematics) to go from these simple abilities to advanced mathematics and theoretical physics.

Three key developments in cognitive science which bear crucially on the maths-physics connection are: (i) embodiment of the mind (our concepts are shaped by the nature of our bodies and brains) (ii) role of the cognitive unconscious (our low level thought processes are not amenable to the conscious mind) (iii) role of conceptual metaphors (our abstract concepts are shaped by our primordial perceptions rooted in the motor-sensory system) \cite{Lakoff}.

Ordinary cognitive mechanisms are shaped by the aforementioned primordial perceptions, which may more formally be classified as: objects and their collections; spatial relations and object distribution in space; time, change and motion; body movements and repeated actions; object manipulation such as rotation and stretching, etc. In technical jargon, there are  four  relevant mechanisms: (i) image schemas (i.e. spatial relations amongst objects);  (ii)  aspect schemas (the important discovery that the same neural structure which is used to control complex motor actions is also used in reasoning \cite{Narayanan}); (iii) conceptual metaphor: mapping from entities in one conceptual domain to another, say from primordial to abstract,  it being a special case of {\it conflation}, which is the simultaneous activation of two distinct areas of the brain, each of which is concerned with a different experience \cite{Johnson}.  Conflation leads to neural connections across domains, resulting in metaphors whereby one domain is conceptualised in terms of another. This is central to understanding how mathematics develops in the brain; (iv) conceptual blends, which combine two distinct cognitive structures \cite{Lakoff}.

As an example, all of arithmetic, including rationals and negative integers, can be deduced, starting from small numbers, using the following four metaphors: Arithmetic as Object Collection, Arithmetic as Object Construction, The Measuring Stick Metaphor, Arithmetic as Motion along a Path.  Algebra is deduced from the metaphor `Essence (essential properties of a thing) is Form (structure)'. Metaphors have also been proposed for understanding irrational numbers, real and complex numbers, limits and the infinitesimal, transfinite numbers and set theory, and infinity. Pathbreaking work has been done in providing a convincing cognitive basis for all of basic mathematics (arithmetic, algebra, geometry, analysis, set theory) \cite{Lakoff}.

Coming to physics, it is reasonable to distinguish between experimental physics [which describes results of experiments in terms of concrete concepts] and theoretical physics, which uses concrete and abstract concepts to give a mathematical formulation of the observed regularity. Theoretical physics could honorably be called a branch of pure mathematics; that branch whose axioms are motivated by experiments on the physical universe. Concrete concepts include quantitative description of mass, force, space and distance, trajectories, time, rotations, currents etc. Force, for instance, could be metaphorically related to the primordial human perception of the muscular exertion in throwing a stone at a prey or a threat. We have hard wired notions of space and motion, and of the flow of psychological time, which is metaphorically concretised to measured laboratory time. It is thus not difficult to accept that the experiments which physicists perform can be described using concrete concepts which draw upon our primordial perceptions. The abstract concepts used in theoretical physics are either the same as those of mathematics (real and complex numbers, calculus and differential equations, Euclidean and Riemannian geometry) or they are expressly invented using conceptual metaphors. For instance, the abstract concept of a field (gravitational field, electromagnetic field) is metaphored from `in an event there is a cause and an influence' and the concept `there can be no action at a distance'. Essentially all of theoretical physics builds on the theme of cause and influence, hard wired into our brains because of the perceived flow of time: (i) how much influence does a source produce at a given location (field equation) and (ii) how does a test object move under this influence (equation of motion). A conceptual metaphor could be: a hunter lights a fire and judges how warm it feels at some distance (field equation) and decides how he should change his position with respect to the fire (equation of motion). 

As an example,  we argue for the cognitive basis of the laws which determine that the orbit of Mars around the sun is an ellipse: a combination of Newton's second law of motion, and his law of gravitation:
\begin{equation}
m_i\frac{d^2{\bf r} }{dt^2} = {\bf F} = -G\frac{Mm_g}{r^3}{\bf {r}}
\end{equation}
Conceptually, it is assumed that Mars, possessed with a heaviness $m_i$, is in orbit around the sun (having a heaviness $M$), and its motion is influenced by the sun (which exerts a force diminishing with distance from the sun). No mathematics so far, except in numerical quantification of mass (which is elementary arithmetic, abstracted from innate arithmetic). Only primordial and concrete experimental concepts such as heaviness, motion, force and distance have been introduced. Next, mathematics is introduced by way of the second law, which encodes the experimentally verified inter-relation between the concrete concepts of mass and force, and the abstract entity from calculus (acceleration as the second time derivative of position). The second equality, the force law of gravitation, is motivated, amongst other things, by the necessity to deduce Kepler's empirical inference that the orbit is an ellipse. The equality of the gravitational mass $m_g$ of Mars (that abstract property which causes it to be influenced by the sun) with its heaviness $m_i$ is another great conceptual leap, motivated by terrestrial experiments of Galileo and others. We clearly see in this example how primordial and concrete concepts amongst which a regularity is observed empirically, are mathematically inter-related by the physicist in his/her brain, using additional abstract concepts, to arrive at a law of nature, using mathematics already known or expressly invented for the job at hand. 

Similar cognitive structure applies in the rest of theoretical physics. [A more advanced example is presented in the Technical Endnotes]. The regularity is in the physical world; its mathematical description is a deliberate fit produced in the brain, using known human mathematics, which is based on known cognitive mechanisms. It is tempting but erroneous to conclude that the beautiful mathematical description is resident {\it in} the physical world out there. Cognition draws upon the physical world to invent the precise, logical  and stable human language of  mathematics. Human mathematics is in turn used by the mind to give a precise description of  the observed  regularity of the physical universe. Such a connection is not mysterious. Rather, it is inevitable. 

%\end{document}

\newpage

\centerline{\bf TECHNICAL ENDNOTES}

\medskip

\centerline{\bf Quantum Theory and Noncommutative Geometry: A Conjecture}

\medskip

\noindent One of the central experimental inputs for quantum theory is the relation $E=\hbar\omega$ between energy and frequency, valid both for massless particles (such as photons) and massive particles (such as electrons). This relation makes it mandatory that the quantum state of a quantum particle, say the electron, must be described by a complex number. This is because the energy $E$ of the particle is necessarily positive, and to a traveling quantum wave described by say
\begin{equation}
\psi\sim \exp^{-i\omega t + ikz}
\end{equation}
one cannot add the complex conjugate, because that will correspond to a negative frequency (and hence negative energy) part, which is disallowed. Thus complex numbers must necessarily enter quantum mechanics if we are to allow for wavelike phenomena, and also allow only for positive frequencies as required by positivity of energy for free particles.

We once again see that first come the experiment on photo-electric effect and the Davisson-Germer experiment establishing the wave nature of massive particles; and then comes the conceptual leap relating energy to frequency. This forces on us the need to describe the state differently, using not real but complex numbers. Cognitive studies trace the origin of complex numbers to the metaphor `Numbers are points on a line' \cite{Lakoff}. Just as a negative integer can be obtained from a positive one by a $180^{o}$ anticlockwise rotation on the number line, the imaginary number $\sqrt{-1}$ is obtained by a $90^{o}$ rotation. In this way, an abstract concept such as complex number has been fitted (in the brain) to describe the observed regularity of the quantum energy-frequency relation, and the positivity of energy.

We now dwell on the all too human (as opposed to Platonic) nature of the maths-physics connection in quantum theory. There are four oddities in this connection, bordering on the illogical, which suggest that quantum theory might be incomplete. The first is what we refer to as the dependence on the so-called classical measuring apparatus. We know from text-books that `when a quantum system interacts with a classical measuring apparatus, the quantum state collapses to one of the eigenstates of the measured observable'. Besides the fact that an entity such as a classical measuring apparatus is vaguely defined, the oddity is that in order to make sense of quantum experiments, the theory has to rely on its own 
limit, i.e. classical mechanics \cite{Bell, LandauLifshitz}. This is unsatisfactory: which came first - quantum theory, or its classical limit? The second oddity is that even though the Schr\"{o}dinger equation is deterministic, the outcomes of a measurement are random, occurring with one or the other probability, given by the Born rule. The initial state is known precisely; there is no initial sampling space, yet the outcomes are probabilistic. This is an extremely unusual state of affairs in physics, the only one of its kind. The mathematical theory of probability is made to stand on its head, so to say! The third oddity is that the collapse of the wave function is instantaneous, so that some kind of acausal influence is felt over spacelike separations, suggesting some tension with special relativity.  The fourth oddity is that quantum theory depends on a classical time for describing evolution, which again is unsatisfactory, for how would one formulate quantum theory if there were no background classical space-time?

Despite these oddities, we physicists feel reluctant to modify quantum theory, for the theory is extremely successful, and is not contradicted by any experiment to date. To us, this shows the frailty of the physics - maths connection: even when the link is not fully logical, we are willing to live with it. Once again, experiments reign supreme in physics.

Nonetheless, we venture to ask, in the spirit in which general relativity was developed, how one might improve this connection, motivated on the one hand by ongoing experiments, and on the other hand by the search for a deeper relation with mathematics. Indeed there exists a well-defined phenomenological modification of non-relativistic quantum theory, known as Continuous Spontaneous Localization [CSL] which explains away the first two oddities, and which is being vigorously tested by ongoing laboratory experiments \cite{RMP:2012, Bera2015}.

The need to do away with classical time in quantum theory seems to suggest interesting new links with mathematics. If one cannot have ordinary classical space-time, then there are reasons to believe that it must be replaced by a noncommutative space-time where space-time coordinates do not commute with each other \cite{Singh:2006}. Such a geometry is a special case of a noncommutative geometry, which is abstracted by first mapping an ordinary geometry to an ordinary commutative algebra, then making the algebra noncommutative, and mapping it back to a new geometry \cite{connesbook}.

Doing so opens up two fascinating  new frontiers. Firstly, it can be argued that the noncommutative space-time can be described by a noncommutative generalisation of special relativity \cite{Lochan-Singh:2011}. Coordinates become matrices/operators, and one has at hand a classical dynamics of matrices, analogous to Adler's Trace Dynamics \cite{Adler:04}. The equilibrium statistical thermodynamics of such a theory can be shown to be a quantum theory without classical time, which in an appropriate limit becomes ordinary quantum theory \cite{Lochan:2012}. The consideration of Brownian motion fluctuations about equilibrium seems to lead to the modified quantum theory described by CSL \cite{Singh:2012}. The oddities of quantum theory seem to go away in a unified manner.

The second remarkable frontier is the connection that noncommutative geometry makes with particle physics and general relativity.  The symmetry group $G$ for the Lagrangian of general relativity and the standard model of particle physics is the semi-direct product of the diffeomorphism group and the group of gauge transformations $SU_3\times SU_2\times U_1$. Now we ask if there is some space $X$ such that $G$ is the group $Diff(X)$. If so, we will have a geometrization of fundamental interactions. However, a no-go theorem forbids that, so long as $X$ is an ordinary (commutative) manifold. On the other hand, if $X$ is a product $M\times F$ of an ordinary manifold $M$ with a finite noncommutative space $F$, such a correspondence  is possible. The algebra of the space $F$ is then determined by the properties and parameters of the standard model \cite{connes, connes2}.

To us, this ongoing  search and endeavour for a better understanding of quantum theory and its relation to space-time structure is one further example of how we humanly abstract from experiments, and then look for appropriate beautiful mathematics to make a better physical theory. It is indeed very hard to envisage that the highly sophisticated mathematics of noncommutative geometry {\it lives} out there in the extremely abstract space $M\times F$, into which our space-time manifold $M$ has been subsumed.  It seems much simpler to accept that the physical laws determined by such a geometry are a product of our mind and live in the human brain.

\bigskip

%\newpage

%\newpage

\centerline{\bf REFERENCES}

\bibliography{biblioqmts3}

\def\polhk#1{\setbox0=\hbox{#1}{\ooalign{\hidewidth
  \lower1.5ex\hbox{`}\hidewidth\crcr\unhbox0}}} \def\cprime{$'$}
  \def\cprime{$'$}
\begin{thebibliography}{30}%
\makeatletter
\providecommand \@ifxundefined [1]{%
 \@ifx{#1\undefined}
}%
\providecommand \@ifnum [1]{%
 \ifnum #1\expandafter \@firstoftwo
 \else \expandafter \@secondoftwo
 \fi
}%
\providecommand \@ifx [1]{%
 \ifx #1\expandafter \@firstoftwo
 \else \expandafter \@secondoftwo
 \fi
}%
\providecommand \natexlab [1]{#1}%
\providecommand \enquote  [1]{``#1''}%
\providecommand \bibnamefont  [1]{#1}%
\providecommand \bibfnamefont [1]{#1}%
\providecommand \citenamefont [1]{#1}%
\providecommand \href@noop [0]{\@secondoftwo}%
\providecommand \href [0]{\begingroup \@sanitize@url \@href}%
\providecommand \@href[1]{\@@startlink{#1}\@@href}%
\providecommand \@@href[1]{\endgroup#1\@@endlink}%
\providecommand \@sanitize@url [0]{\catcode `\\12\catcode `\$12\catcode
  `\&12\catcode `\#12\catcode `\^12\catcode `\_12\catcode `\%12\relax}%
\providecommand \@@startlink[1]{}%
\providecommand \@@endlink[0]{}%
\providecommand \url  [0]{\begingroup\@sanitize@url \@url }%
\providecommand \@url [1]{\endgroup\@href {#1}{\urlprefix }}%
\providecommand \urlprefix  [0]{URL }%
\providecommand \Eprint [0]{\href }%
\providecommand \doibase [0]{http://dx.doi.org/}%
\providecommand \selectlanguage [0]{\@gobble}%
\providecommand \bibinfo  [0]{\@secondoftwo}%
\providecommand \bibfield  [0]{\@secondoftwo}%
\providecommand \translation [1]{[#1]}%
\providecommand \BibitemOpen [0]{}%
\providecommand \bibitemStop [0]{}%
\providecommand \bibitemNoStop [0]{.\EOS\space}%
\providecommand \EOS [0]{\spacefactor3000\relax}%
\providecommand \BibitemShut  [1]{\csname bibitem#1\endcsname}%
\let\auto@bib@innerbib\@empty
%</preamble>
\bibitem [{\citenamefont {Courant}\ and\ \citenamefont
  {Robbins}(1996)}]{Courant}%
  \BibitemOpen
  \bibfield  {author} {\bibinfo {author} {\bibfnamefont {R.}~\bibnamefont
  {Courant}}\ and\ \bibinfo {author} {\bibfnamefont {H.}~\bibnamefont
  {Robbins}},\ }\href@noop {} {\emph {\bibinfo {title} {What is
  Mathematics?}}}\ (\bibinfo  {publisher} {Oxford University Press},\ \bibinfo
  {year} {1996})\BibitemShut {NoStop}%
\bibitem [{\citenamefont {Koestler}(1990)}]{Koestler}%
  \BibitemOpen
  \bibfield  {author} {\bibinfo {author} {\bibfnamefont {A.}~\bibnamefont
  {Koestler}},\ }\href@noop {} {\emph {\bibinfo {title} {The Sleepwalkers: A
  History of Man's Changing Vision of the Universe}}}\ (\bibinfo  {publisher}
  {Penguin Books},\ \bibinfo {year} {1990})\BibitemShut {NoStop}%
\bibitem [{\citenamefont {Misner}\ \emph {et~al.}(1973)\citenamefont {Misner},
  \citenamefont {Thorne},\ and\ \citenamefont {Wheeler}}]{MTW}%
  \BibitemOpen
  \bibfield  {author} {\bibinfo {author} {\bibfnamefont {C.~W.}\ \bibnamefont
  {Misner}}, \bibinfo {author} {\bibfnamefont {K.}~\bibnamefont {Thorne}}, \
  and\ \bibinfo {author} {\bibfnamefont {J.~A.}\ \bibnamefont {Wheeler}},\
  }\href@noop {} {\emph {\bibinfo {title} {Gravitation}}}\ (\bibinfo
  {publisher} {W. H. Freeman},\ \bibinfo {year} {1973})\BibitemShut {NoStop}%
\bibitem [{\citenamefont {Ruelle}(2007)}]{Ruelle}%
  \BibitemOpen
  \bibfield  {author} {\bibinfo {author} {\bibfnamefont {D.}~\bibnamefont
  {Ruelle}},\ }\href@noop {} {\emph {\bibinfo {title} {The Mathematician's
  Brain}}}\ (\bibinfo  {publisher} {Princeton University Press},\ \bibinfo
  {year} {2007})\BibitemShut {NoStop}%
\bibitem [{\citenamefont {Appel}\ and\ \citenamefont
  {Wolfgang}(1977)}]{FourColor}%
  \BibitemOpen
  \bibfield  {author} {\bibinfo {author} {\bibfnamefont {K.}~\bibnamefont
  {Appel}}\ and\ \bibinfo {author} {\bibfnamefont {H.}~\bibnamefont
  {Wolfgang}},\ }\href@noop {} {\bibfield  {journal} {\bibinfo  {journal}
  {Illinois Journal of Mathematics}\ }\textbf {\bibinfo {volume} {21}},\
  \bibinfo {pages} {429} (\bibinfo {year} {1977})}\BibitemShut {NoStop}%
\bibitem [{\citenamefont {Faltings}(1983)}]{Falting}%
  \BibitemOpen
  \bibfield  {author} {\bibinfo {author} {\bibfnamefont {G.}~\bibnamefont
  {Faltings}},\ }\href@noop {} {\bibfield  {journal} {\bibinfo  {journal}
  {Inventiones Mathematicae}\ }\textbf {\bibinfo {volume} {73}},\ \bibinfo
  {pages} {349} (\bibinfo {year} {1983})}\BibitemShut {NoStop}%
\bibitem [{\citenamefont {Jones}(1985)}]{Jones}%
  \BibitemOpen
  \bibfield  {author} {\bibinfo {author} {\bibfnamefont {V.}~\bibnamefont
  {Jones}},\ }\href@noop {} {\bibfield  {journal} {\bibinfo  {journal} {Bull.
  Amer. Math. Soc. (N.S.)}\ }\textbf {\bibinfo {volume} {12}},\ \bibinfo
  {pages} {103} (\bibinfo {year} {1985})}\BibitemShut {NoStop}%
\bibitem [{\citenamefont {Witten}(1989)}]{Witten}%
  \BibitemOpen
  \bibfield  {author} {\bibinfo {author} {\bibfnamefont {E.}~\bibnamefont
  {Witten}},\ }\href@noop {} {\bibfield  {journal} {\bibinfo  {journal}
  {Commun. Math. Phys.}\ }\textbf {\bibinfo {volume} {121}},\ \bibinfo {pages}
  {351} (\bibinfo {year} {1989})}\BibitemShut {NoStop}%
\bibitem [{\citenamefont {Dehaene}(2011)}]{Dehaene}%
  \BibitemOpen
  \bibfield  {author} {\bibinfo {author} {\bibfnamefont {S.}~\bibnamefont
  {Dehaene}},\ }\href@noop {} {\emph {\bibinfo {title} {The Number Sense: How
  the Mind Creates Mathematics}}}\ (\bibinfo  {publisher} {Oxford University
  Press},\ \bibinfo {year} {2011})\BibitemShut {NoStop}%
\bibitem [{\citenamefont {Antell}\ and\ \citenamefont
  {Keating}(1983)}]{Antell}%
  \BibitemOpen
  \bibfield  {author} {\bibinfo {author} {\bibfnamefont {S.~E.}\ \bibnamefont
  {Antell}}\ and\ \bibinfo {author} {\bibfnamefont {D.~P.}\ \bibnamefont
  {Keating}},\ }\href@noop {} {\bibfield  {journal} {\bibinfo  {journal} {Child
  Development}\ }\textbf {\bibinfo {volume} {54}},\ \bibinfo {pages} {695}
  (\bibinfo {year} {1983})}\BibitemShut {NoStop}%
\bibitem [{\citenamefont {Wynn}(1992)}]{Wynn}%
  \BibitemOpen
  \bibfield  {author} {\bibinfo {author} {\bibfnamefont {K.}~\bibnamefont
  {Wynn}},\ }\href@noop {} {\bibfield  {journal} {\bibinfo  {journal} {Nature}\
  }\textbf {\bibinfo {volume} {358}},\ \bibinfo {pages} {749} (\bibinfo {year}
  {1992})}\BibitemShut {NoStop}%
\bibitem [{\citenamefont {Bijeljac-Babic}\ \emph {et~al.}(1991)\citenamefont
  {Bijeljac-Babic}, \citenamefont {Bertoncini},\ and\ \citenamefont
  {Mehler}}]{Bibelac}%
  \BibitemOpen
  \bibfield  {author} {\bibinfo {author} {\bibfnamefont {R.}~\bibnamefont
  {Bijeljac-Babic}}, \bibinfo {author} {\bibfnamefont {J.}~\bibnamefont
  {Bertoncini}}, \ and\ \bibinfo {author} {\bibfnamefont {J.}~\bibnamefont
  {Mehler}},\ }\href@noop {} {\bibfield  {journal} {\bibinfo  {journal}
  {Developmental Psychology}\ }\textbf {\bibinfo {volume} {29}},\ \bibinfo
  {pages} {711} (\bibinfo {year} {1991})}\BibitemShut {NoStop}%
\bibitem [{\citenamefont {Mechner}\ and\ \citenamefont
  {Gueverekian}(1962)}]{Mechner}%
  \BibitemOpen
  \bibfield  {author} {\bibinfo {author} {\bibfnamefont {F.}~\bibnamefont
  {Mechner}}\ and\ \bibinfo {author} {\bibfnamefont {L.}~\bibnamefont
  {Gueverekian}},\ }\href@noop {} {\bibfield  {journal} {\bibinfo  {journal}
  {Journal of the Experimental Analysis of Behavior}\ }\textbf {\bibinfo
  {volume} {5}},\ \bibinfo {pages} {463} (\bibinfo {year} {1962})}\BibitemShut
  {NoStop}%
\bibitem [{\citenamefont {Church}\ and\ \citenamefont {Meck}(1984)}]{Church}%
  \BibitemOpen
  \bibfield  {author} {\bibinfo {author} {\bibfnamefont {R.~M.}\ \bibnamefont
  {Church}}\ and\ \bibinfo {author} {\bibfnamefont {W.~H.}\ \bibnamefont
  {Meck}},\ }\href@noop {} {\emph {\bibinfo {title} {Animal Cognition}}},\
  edited by\ \bibinfo {editor} {\bibfnamefont {T.~G.}\ \bibnamefont {Bever}}\
  and\ \bibinfo {editor} {\bibfnamefont {H.~S.}\ \bibnamefont {Terrace}}\
  (\bibinfo  {publisher} {Hillsdale, N.J.:Erlbaum},\ \bibinfo {year}
  {1984})\BibitemShut {NoStop}%
\bibitem [{\citenamefont {Hauser}\ \emph {et~al.}(1996)\citenamefont {Hauser},
  \citenamefont {MacNeilage},\ and\ \citenamefont {Ware}}]{Hauser}%
  \BibitemOpen
  \bibfield  {author} {\bibinfo {author} {\bibfnamefont {M.~D.}\ \bibnamefont
  {Hauser}}, \bibinfo {author} {\bibfnamefont {P.}~\bibnamefont {MacNeilage}},
  \ and\ \bibinfo {author} {\bibfnamefont {M.}~\bibnamefont {Ware}},\
  }\href@noop {} {\bibfield  {journal} {\bibinfo  {journal} {Proc. Nat. Aca.
  Sci. USA}\ }\textbf {\bibinfo {volume} {93}},\ \bibinfo {pages} {1514}
  (\bibinfo {year} {1996})}\BibitemShut {NoStop}%
\bibitem [{\citenamefont {Lakoff}\ and\ \citenamefont {Nunez}(2000)}]{Lakoff}%
  \BibitemOpen
  \bibfield  {author} {\bibinfo {author} {\bibfnamefont {G.}~\bibnamefont
  {Lakoff}}\ and\ \bibinfo {author} {\bibfnamefont {R.~E.}\ \bibnamefont
  {Nunez}},\ }\href@noop {} {\emph {\bibinfo {title} {Where Mathematics Comes
  From?}}}\ (\bibinfo  {publisher} {Basic Books},\ \bibinfo {year}
  {2000})\BibitemShut {NoStop}%
\bibitem [{\citenamefont {Narayanan}(1997)}]{Narayanan}%
  \BibitemOpen
  \bibfield  {author} {\bibinfo {author} {\bibfnamefont {S.}~\bibnamefont
  {Narayanan}},\ }\emph {\bibinfo {title} {Embodiment in language
  understanding: sensory motor representations for metaphoric reasoning about
  event descriptions}},\ \href@noop {} {Ph.D. thesis},\ \bibinfo  {school} {UC
  Berkeley} (\bibinfo {year} {1997})\BibitemShut {NoStop}%
\bibitem [{\citenamefont {Johnson}(1997)}]{Johnson}%
  \BibitemOpen
  \bibfield  {author} {\bibinfo {author} {\bibfnamefont {C.}~\bibnamefont
  {Johnson}},\ }\href@noop {} {\emph {\bibinfo {title} {Metaphor vs. conflation
  in the acquisition of polysemy: The case of SEE}}},\ edited by\ \bibinfo
  {editor} {\bibfnamefont {M.~K.}\ \bibnamefont {Hiraga}}, \bibinfo {editor}
  {\bibfnamefont {C.}~\bibnamefont {Sinha}}, \ and\ \bibinfo {editor}
  {\bibfnamefont {S.}~\bibnamefont {Wilcox}}\ (\bibinfo  {publisher}
  {Amsterdam: John Benjamins},\ \bibinfo {year} {1997})\BibitemShut {NoStop}%
\bibitem [{\citenamefont {Bell}(1990)}]{Bell}%
  \BibitemOpen
  \bibfield  {author} {\bibinfo {author} {\bibfnamefont {J.~S.}\ \bibnamefont
  {Bell}},\ }\href@noop {} {\bibfield  {journal} {\bibinfo  {journal} {Physics
  World}\ ,\ \bibinfo {pages} {8, 33}} (\bibinfo {year} {1990})}\BibitemShut
  {NoStop}%
\bibitem [{\citenamefont {Landau}\ and\ \citenamefont
  {Lifshitz}(1965)}]{LandauLifshitz}%
  \BibitemOpen
  \bibfield  {author} {\bibinfo {author} {\bibfnamefont {L.~D.}\ \bibnamefont
  {Landau}}\ and\ \bibinfo {author} {\bibfnamefont {E.~M.}\ \bibnamefont
  {Lifshitz}},\ }\href@noop {} {\emph {\bibinfo {title} {Quantum Mechanics}}}\
  (\bibinfo  {publisher} {Pergamon Press},\ \bibinfo {year} {1965})\BibitemShut
  {NoStop}%
\bibitem [{\citenamefont {Bassi}\ \emph {et~al.}(2013)\citenamefont {Bassi},
  \citenamefont {Lochan}, \citenamefont {Satin}, \citenamefont {Singh},\ and\
  \citenamefont {Ulbricht}}]{RMP:2012}%
  \BibitemOpen
  \bibfield  {author} {\bibinfo {author} {\bibfnamefont {A.}~\bibnamefont
  {Bassi}}, \bibinfo {author} {\bibfnamefont {K.}~\bibnamefont {Lochan}},
  \bibinfo {author} {\bibfnamefont {S.}~\bibnamefont {Satin}}, \bibinfo
  {author} {\bibfnamefont {T.~P.}\ \bibnamefont {Singh}}, \ and\ \bibinfo
  {author} {\bibfnamefont {H.}~\bibnamefont {Ulbricht}},\ }\href@noop {}
  {\bibfield  {journal} {\bibinfo  {journal} {Rev. Mod. Phys.}\ }\textbf
  {\bibinfo {volume} {85}},\ \bibinfo {pages} {471} (\bibinfo {year}
  {2013})}\BibitemShut {NoStop}%
\bibitem [{\citenamefont {Bera}\ \emph {et~al.}(2015)\citenamefont {Bera},
  \citenamefont {Motwani}, \citenamefont {Singh},\ and\ \citenamefont
  {Ulbricht}}]{Bera2015}%
  \BibitemOpen
  \bibfield  {author} {\bibinfo {author} {\bibfnamefont {S.}~\bibnamefont
  {Bera}}, \bibinfo {author} {\bibfnamefont {B.}~\bibnamefont {Motwani}},
  \bibinfo {author} {\bibfnamefont {T.~P.}\ \bibnamefont {Singh}}, \ and\
  \bibinfo {author} {\bibfnamefont {H.}~\bibnamefont {Ulbricht}},\ }\href@noop
  {} {\bibfield  {journal} {\bibinfo  {journal} {Scientific Reports}\ }\textbf
  {\bibinfo {volume} {5}},\ \bibinfo {pages} {7664} (\bibinfo {year}
  {2015})}\BibitemShut {NoStop}%
\bibitem [{\citenamefont {Singh}(2006)}]{Singh:2006}%
  \BibitemOpen
  \bibfield  {author} {\bibinfo {author} {\bibfnamefont {T.~P.}\ \bibnamefont
  {Singh}},\ }\href@noop {} {\bibfield  {journal} {\bibinfo  {journal} {Bulg.
  J. Phys.}\ }\textbf {\bibinfo {volume} {33}},\ \bibinfo {pages} {217}
  (\bibinfo {year} {2006})}\BibitemShut {NoStop}%
\bibitem [{\citenamefont {Connes}(1995)}]{connesbook}%
  \BibitemOpen
  \bibfield  {author} {\bibinfo {author} {\bibfnamefont {A.}~\bibnamefont
  {Connes}},\ }\href@noop {} {\emph {\bibinfo {title} {Noncommutative
  Geometry}}}\ (\bibinfo  {publisher} {Academic Press Inc},\ \bibinfo {year}
  {1995})\BibitemShut {NoStop}%
\bibitem [{\citenamefont {Lochan}\ and\ \citenamefont
  {Singh}(2011)}]{Lochan-Singh:2011}%
  \BibitemOpen
  \bibfield  {author} {\bibinfo {author} {\bibfnamefont {K.}~\bibnamefont
  {Lochan}}\ and\ \bibinfo {author} {\bibfnamefont {T.~P.}\ \bibnamefont
  {Singh}},\ }\href@noop {} {\bibfield  {journal} {\bibinfo  {journal} {Phys.
  Lett. A}\ }\textbf {\bibinfo {volume} {375}},\ \bibinfo {pages} {3747}
  (\bibinfo {year} {2011})}\BibitemShut {NoStop}%
\bibitem [{\citenamefont {Adler}(2004)}]{Adler:04}%
  \BibitemOpen
  \bibfield  {author} {\bibinfo {author} {\bibfnamefont {S.~L.}\ \bibnamefont
  {Adler}},\ }\href@noop {} {\emph {\bibinfo {title} {Quantum theory as an
  emergent phenomenon}}}\ (\bibinfo  {publisher} {Cambridge University Press},\
  \bibinfo {address} {Cambridge},\ \bibinfo {year} {2004})\ pp.\ \bibinfo
  {pages} {xii+225}\BibitemShut {NoStop}%
\bibitem [{\citenamefont {Lochan}\ \emph {et~al.}(2012)\citenamefont {Lochan},
  \citenamefont {Satin},\ and\ \citenamefont {Singh}}]{Lochan:2012}%
  \BibitemOpen
  \bibfield  {author} {\bibinfo {author} {\bibfnamefont {K.}~\bibnamefont
  {Lochan}}, \bibinfo {author} {\bibfnamefont {S.}~\bibnamefont {Satin}}, \
  and\ \bibinfo {author} {\bibfnamefont {T.~P.}\ \bibnamefont {Singh}},\
  }\href@noop {} {\bibfield  {journal} {\bibinfo  {journal} {Found. Phys.}\
  }\textbf {\bibinfo {volume} {42}},\ \bibinfo {pages} {1556} (\bibinfo {year}
  {2012})}\BibitemShut {NoStop}%
\bibitem [{\citenamefont {Singh}(2013)}]{Singh:2012}%
  \BibitemOpen
  \bibfield  {author} {\bibinfo {author} {\bibfnamefont {T.~P.}\ \bibnamefont
  {Singh}},\ }in\ \href@noop {} {\emph {\bibinfo {booktitle} {The Forgotten
  Present}}},\ \bibinfo {editor} {edited by\ \bibinfo {editor} {\bibfnamefont
  {T.}~\bibnamefont {Filk}}\ and\ \bibinfo {editor} {\bibfnamefont
  {A.}~\bibnamefont {von Muller~(arXiv:1210.8110)}}}\ (\bibinfo  {publisher}
  {Springer: Berlin-Heidelberg},\ \bibinfo {year} {2013})\BibitemShut {NoStop}%
\bibitem [{\citenamefont {Connes}(2000)}]{connes}%
  \BibitemOpen
  \bibfield  {author} {\bibinfo {author} {\bibfnamefont {A.}~\bibnamefont
  {Connes}},\ }\href@noop {} {\ \textbf {\bibinfo {volume}
  {arXiv:math/0011193}} (\bibinfo {year} {2000})}\BibitemShut {NoStop}%
\bibitem [{\citenamefont {Chamseddine}\ and\ \citenamefont
  {Connes}(2010)}]{connes2}%
  \BibitemOpen
  \bibfield  {author} {\bibinfo {author} {\bibfnamefont {A.~H.}\ \bibnamefont
  {Chamseddine}}\ and\ \bibinfo {author} {\bibfnamefont {A.}~\bibnamefont
  {Connes}},\ }\href@noop {} {\bibfield  {journal} {\bibinfo  {journal}
  {Fortsch. Phys.}\ }\textbf {\bibinfo {volume} {58}},\ \bibinfo {pages} {553}
  (\bibinfo {year} {2010})}\BibitemShut {NoStop}%
\end{thebibliography}%

\end{document}